\documentclass[a4paper]{article}

\usepackage{INTERSPEECH2021}
\usepackage{nameref}
\usepackage{hyperref}
\usepackage{soul}
\usepackage{color}
\usepackage{listings}
\usepackage[symbol]{footmisc}
\usepackage{dsfont}
\usepackage{amsmath}
\usepackage{mathtools}
\usepackage{centernot}
\usepackage{multirow}
\usepackage{makecell}

\DeclarePairedDelimiter\abs{\lvert}{\rvert}%
\makeatletter
\let\oldabs\abs
\def\abs{\@ifstar{\oldabs}{\oldabs*}}

\newcommand{\norm}[1]{\left\lVert#1\right\rVert}

\newcommand{\parn}[1]{\left(#1\right)}

\newlength{\NOTskip}

\title{StarGANv2-VC: A Diverse, Unsupervised, Non-parallel Framework for Natural-Sounding Voice Conversion}

\name{Yinghao Aaron Li, Ali Zare, Nima Mesgarani}
\address{
  Department of Electrical Engineering, Columbia University, USA}
\email{yl4579@columbia.edu, az2584@columbia.edu, nima@ee.columbia.edu}

\begin{document}

\maketitle
\begin{abstract}
    We present an unsupervised non-parallel many-to-many voice conversion (VC) method using a generative adversarial network (GAN) called StarGAN v2. Using a combination of adversarial source classifier loss and perceptual loss, our model significantly outperforms previous VC models. Although our model is trained only with 20 English speakers, it generalizes to a variety of voice conversion tasks, such as any-to-many, cross-lingual, and singing conversion. Using a style encoder, our framework can also convert plain reading speech into stylistic speech, such as emotional and falsetto speech. Subjective and objective evaluation experiments on a non-parallel many-to-many voice conversion task revealed that our model produces natural sounding voices, close to the sound quality of state-of-the-art text-to-speech (TTS) based voice conversion methods without the need for text labels. Moreover, our model is completely convolutional and with a faster-than-real-time vocoder such as Parallel WaveGAN can perform real-time voice conversion.

\end{abstract}

\section{Introduction}
    Voice conversion (VC) is a technique for converting one speaker's voice identity into another while preserving linguistic content. This technique has various applications, such as movie dubbing, language learning by cross-language conversion, speaking assistance, and singing conversion. However, most voice conversion methods require parallel utterances to achieve high-quality natural conversion results, which strongly limits the conditions where this technique can be applied. 
    
    Recent work in non-parallel voice conversion using deep neural network models can mainly be divided into three categories: auto-encoder-based approach, TTS-based approach, and GAN-based approach. Auto-encoder approach, such as in \cite{qian2019autovc, ding2019group, huang2020unsupervised, qian2020f0}, seeks to encode speaker-independent information from input audio by training models with proper constraints. This approach requires carefully designed constraints to remove speaker-dependent information, and the converted speech quality depends on how much linguistic information can be retrieved from the latent space. On the other hand, GAN-based approaches, such as CycleGAN-VC3\cite{kaneko2020cyclegan} and StarGAN-VC2\cite{kaneko2019stargan} do not constrain the encoder, instead, they use a discriminator that teaches the decoder to generate speech that sounds like the target speaker. Since there is no guarantee that the discriminator will learn meaningful features from the real data, this approach often suffers from problems such as dissimilarity between converted and target speech, or distortions in voices of the generated speech. Unlike the previous two methods, TTS-based approaches like Cotatron \cite{park2020cotatron}, AttS2S-VC \cite{tanaka2019atts2s} and VTN \cite{huang2019voice} take advantage of text labels and synthesizes speech directly by extracting aligned linguistic features from the input speech. This ensures that the converted speaker identity is the same as the target speaker identity. However, this approach requires text labels, which are not often available at hand. 
    
    Here, we present a new method for unsupervised non-parallel many-to-many voice conversion using recently proposed GAN architecture for image style transfer, StarGAN v2 \cite{choi2020stargan}. Our framework produces natural-sounding speech and significantly outperforms the previous state-of-art method, AUTO-VC \cite{qian2019autovc}, in terms of both naturalness and speaker similarity, approaching the TTS based approaches such as VTN \cite{huang2019voice} as reported in Voice Conversion Challenge 2020 (VCC2020) \cite{zhao2020voice} . Besides, our model can generalize to a variety of voice conversion tasks, including any-to-many conversion, cross-language conversion, and singing conversion, even though it was trained only on monolingual speech data with limited numbers of speakers. Furthermore, when trained on a corpus with diverse speech styles, our model shows the ability to convert into stylistic speech, such as converting a plain reading voice into an emotive acting voice and convert a chest voice into a falsetto voice.
    
    We have multiple contributions in this work: (i) applying StarGAN v2 to voice conversion, which enables converting from plain speech into speech with a diversity of styles, (ii) introduce a novel adversarial source classifier loss that greatly improves the similarity in terms of speaker identity between the converted speech and target speech, and (iii) the first voice conversion framework, as far as we know, that employs perceptual losses using both automatic speech recognition (ASR) network and fundamental frequency (F0) extraction network.
    
\section{Method}

\begin{figure*}[t]
  \centering
  \includegraphics[width=\linewidth]{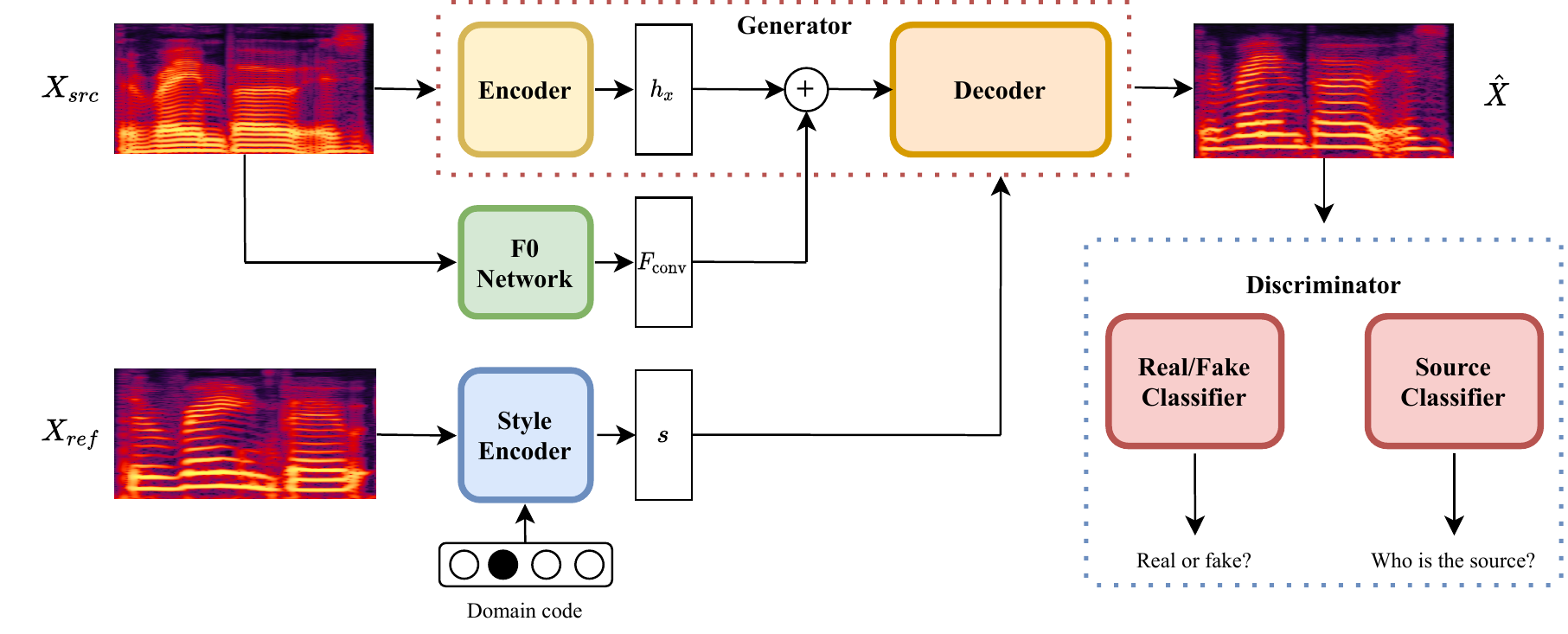}
  \caption{StarGANv2-VC framework with style encoder. $\bm{X_{src}}$ is the source input, $\bm{X_{ref}}$ is the reference input that contains the style information, and $\bm{\hat{X}}$ represents the converted mel-spectrogram. $h_x$, $F_\text{conv}$ and $s$ denote the latent feature of the source, the F0 feature from convolutional layers of the source, and the style code of the reference in the target domain, respectively. $h_x$ and $h_F0$ are concatenated by channel as the input to the decoder, and $h_{sty}$ is injected into the decoder by the adaptive instance normalization (AdaIN) \cite{huang2017arbitrary}. Two classifiers form the discriminators that determine whether a generated sample is real or fake and who the source speaker of $\bm{\hat{X}}$ is. In another scheme where the style encoder is replaced with the mapping network, the reference mel-spectrogram $\bm{X_{ref}}$ is not needed. }
  \label{fig:starganv2vc}
\end{figure*}
    
    \subsection{StarGANv2-VC} 
    StarGAN v2 \cite{choi2020stargan} uses a single discriminator and generator to generate diverse images in each domain with the domain-specific style vectors from either the style encoder or the mapping network. We have adopted the same architecture to voice conversion, treated each speaker as an individual domain, and added a pre-trained joint detection and classification (JDC) F0 extraction network \cite{kum2019joint} to achieve F0-consistent conversion. An overview of our framework is shown in Figure~\ref{fig:starganv2vc}.
    
   \noindent\textbf{Generator.} The generator $G$ converts an input mel-spectrogram $\bm{X_{src}}$ into $G(\bm{X_{src}}, h_{sty}, h_{f0})$ that reflects the style in $h_{sty}$, which is given either by the mapping network or the style encoder, and the fundamental frequency in $h_{f0}$, which is provided by the convolution layers in the F0 extraction network $F$. 
   
   \noindent\textbf{F0 network.} The F0 extraction network $F$ is a pre-trained JDC network \cite{kum2019joint} that extracts the fundamental frequency from an input mel-spectrogram. The JDC  network has convolutional layers followed by BLSTM units. We only use the convolutional output $F_{\text{conv}}(\bm{X})$ for $\bm{X} \in \mathcal{X}$ as the input features. 
   
   \noindent\textbf{Mapping network.} The mapping network $M$ generates a style vector $h_{M} = M(\bm{z}, y)$ with a random latent code $\bm{z} \in \mathcal{Z}$ in a domain $y \in \mathcal{Y}$. The latent code is sampled from a Gaussian distribution to provide diverse style representations in all domains. The style vector representation is shared for all domains until the last layer, where a domain-specific projection is applied to the shared representation.
   
   \noindent\textbf{Style encoder.} Given a reference  mel-spectrogram $\bm{X_{ref}}$, the style encoder $S$ extracts the style code $h_{sty} = S(\bm{X_{ref}}, y)$ in the domain $y \in \mathcal{Y}$. Similar to the mapping network $M$, $S$ first processes an input through shared layers across all domains. A domain-specific projection then maps the shared features into a domain-specific style code. 
   
\begin{figure*}[t]
  \centering
  \includegraphics[width=\linewidth]{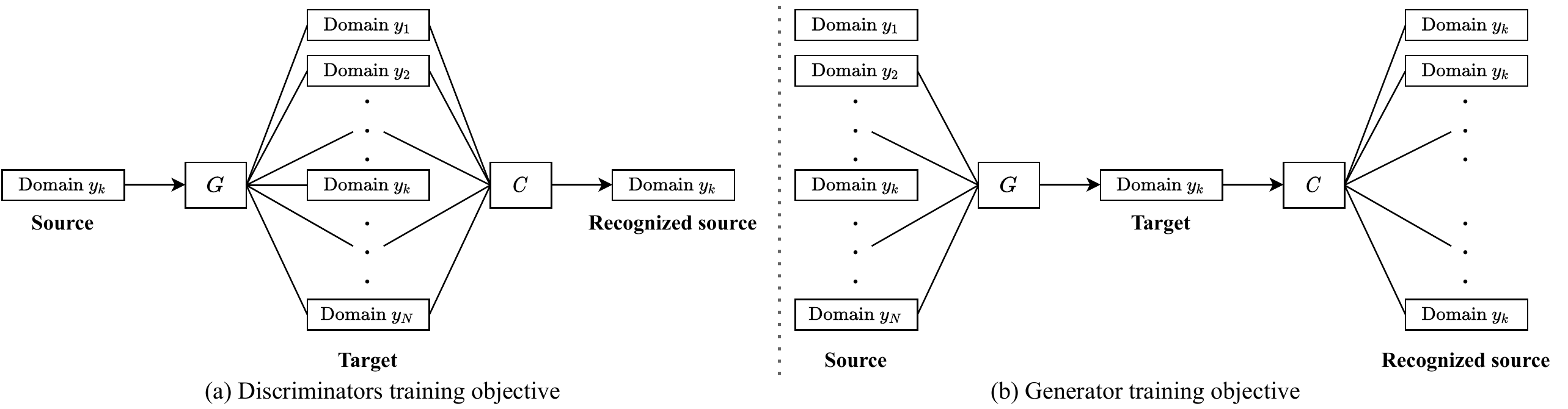}
  \caption{Training schemes of adversarial source classifier for a domain $y_k$. The case $y_{trg} = y_{src}$ is omitted to prevent amplification of artifacts from the source classifier. \textbf{(a)} When training the discriminators, the weights of the generator $G$ are fixed, and the source classifier $C$ is trained to determine the original domain $y_k$ of the converted samples, regardless of the target domains. \textbf{(b)} When training the generator, the weights of the source classifier $C$ are fixed, and the generator $G$ is trained to make $C$ classify all generated samples as being converted from the target domain $y_k$, regardless of the actual domains of the source.}
  \label{fig:source_classifier}
\end{figure*}

   \noindent\textbf{Discriminators.} The discriminator $D$ in \cite{choi2020stargan} has shared layers that learns the common features between real and fake samples in all domains, followed by a domain-specific binary classifier that classifies whether a sample is real in each domain $y \in \mathcal{Y}$. However, since the domain-specific classifier consists of only one convolutional layer, it may fail to capture important aspects of domain-specific features such as the pronunciations of a speaker. To address this problem, we introduce an additional classifier $C$ with the same architecture as $D$ that learns the original domain of converted samples. By learning what features elude the input domain even after conversion, the classifier can provide feedback about features invariant to the generator yet characteristic to the original domain, upon which the generator should improve to generate a more similar sample in the target domain. A more detailed illustration is given in Figure~\ref{fig:source_classifier}.

    \subsection{Training Objectives}
    The aim of StarGANv2-VC is to learn a mapping $G:  \mathcal{X}_{y_{src}} \rightarrow \mathcal{X}_{y_{trg}}$ that converts a sample $\bm{X}\in \mathcal{X}_{y_{src}}$ from the source domain $y_{src} \in \mathcal{Y}$ to a sample $\bm{\hat{X}} \in \mathcal{X}_{y_{trg}}$ in the target domain $y_{trg} \in \mathcal{Y}$ without parallel data. 
    
    During training, we sample a target domain $y_{trg} \in \mathcal{Y}$ and a style code $s \in \mathcal{S}_{y_{trg}}$ randomly via either mapping network where $s = M(z, y_{trg})$ with a latent code $z \in \mathcal{Z}$, or style encoder where $s = S(\bm{X_{ref}}, y_{trg})$ with a reference input $\bm{X_{ref}} \in \mathcal{X}$. Given a mel-spectrogram $\bm{X}\in \mathcal{X}_{y_{src}}$, the source domain $y_{src} \in \mathcal{Y}$ and the target domain $y_{trg} \in \mathcal{Y}$, we train our model with the following loss functions.
    
    \noindent\textbf{Adversarial loss.} The generator takes an input mel-spectrogram $\bm{X}$ and a style vector $s$ and learns to generate a new mel-spectrogram $G(\bm{X}, s)$ via the adversarial loss
    \begin{equation}
    \begin{aligned}    
      \mathcal{L}_{adv} =\text{ } &\mathbb{E}_{\bm{X}, y_{src}}\left[\log D(\bm{X}, y_{src})\right] +\\
      &\mathbb{E}_{\bm{X}, y_{trg}, s}\left[\log\parn{1 - D(G(\bm{X}, s), y_{trg})}\right]
    \end{aligned}
      \label{eq1}
    \end{equation}
    where $D(\cdot, y)$ denotes the output of real/fake classifier for the domain $y \in \mathcal{Y}$. 
    
    \noindent\textbf{Adversarial source classifier loss.} We use an additional adversarial loss function with the source classifier $C$ (see Figure~\ref{fig:source_classifier})
    \begin{equation}
      \mathcal{L}_{advcls} = \mathbb{E}_{\bm{X}, y_{trg}, s}\left[\text{CE}(C(G(\bm{X}, s)), y_{trg})\right]
      \label{eq2}
    \end{equation}
    where $\text{CE}(\cdot)$ denotes the cross-entropy loss function.
    
    \noindent\textbf{Style reconstruction loss.} We use the style reconstruction loss to ensure that the style code can be reconstructed from the generated samples.
    \begin{equation}
      \mathcal{L}_{sty} = \mathbb{E}_{\bm{X}, y_{trg}, s}\left[\norm{s - S(G(\bm{X}, s), y_{trg})}_1\right]
      \label{eq3}
    \end{equation}
    
    \noindent\textbf{Style diversification loss.} The style diversification loss is maximized to enforce the generator to generate different samples with different style codes. In addition to maximizing the mean absolute error (MAE) between generated samples, we also maximize MAE of the F0 features between samples generated with different style codes
    \begin{equation}
    \begin{aligned}  
      &\mathcal{L}_{ds} =\text{ } \mathbb{E}_{\bm{X}, s_1, s_2, y_{trg}}\left[\norm{G(\bm{X}, s_1) - G(\bm{X}, s_2))}_1\right] + 
      \\ &\mathbb{E}_{\bm{X}, s_1, s_2, y_{trg}}\left[\norm{F_{\text{conv}}(G(\bm{X}, s_1)) - F_{\text{conv}}(G(\bm{X}, s_2)))}_1\right]
    \end{aligned}
      \label{eq4}
    \end{equation}
    where $s_1, s_2 \in \mathcal{S}_{y_{trg}}$ are two randomly sampled style codes from domain $y_{trg} \in \mathcal{Y}$ and $F_{\text{conv}}(\cdot)$ is the output of convolutional layers of F0 network $F$ . 
    
    \noindent\textbf{F0 consistency loss.} 
    To produce F0-consistent results, we add an F0-consistent loss with the normalized F0 curve provided by F0 network $F$. For an input mel-spectrogram $\bm{X}$, $F(\bm{X})$ provides the absolute F0 value in Hertz for each frame of $\bm{X}$. Since male and female speakers have different average F0, we normalize the absolute F0 values $F(\bm{X})$ by its temporal mean, denoted by $\hat{F}(\bm{X}) = \frac{F(\bm{X})}{\norm{F(\bm{X})}_1}$. The F0 consistency loss is thus
    \begin{equation}
      \mathcal{L}_{f0} = \mathbb{E}_{\bm{X}, s}\left[\norm{\hat{F}(\bm{X}) - \hat{F}(G(\bm{X}, s))}_1\right]
      \label{eq5}
    \end{equation}
    
    \noindent\textbf{Speech consistency loss.} 
    To ensure that the converted speech has the same linguistic content as the source, we employ a speech consistency loss using convolutional features from a pre-trained joint CTC-attention VGG-BLSTM network \cite{kim2017joint} given in Espnet toolkit
    \footnote{\url{https://github.com/espnet/espnet}}\cite{watanabe2018espnet}. Similar to \cite{polyak2020unsupervised}, we use the output from the intermediate layer before the LSTM layers as the linguistic feature, denoted by $h_{asr}(\cdot)$. The speech consistency loss is defined as
    \begin{equation}
      \mathcal{L}_{asr} = \mathbb{E}_{\bm{X}, s}\left[\norm{h_{asr}(\bm{X}) - h_{asr}(G(\bm{X}, s))}_1\right]
      \label{eq6}
    \end{equation}
    
    \noindent\textbf{Norm consistency loss.} We use the norm consistency loss to preserve the speech/silence intervals of generated samples. We use the absolute column-sum norm for a mel-spectrogram $\bm{X}$ with $N$ mels and $T$ frames at the $t^{th}$ frame, defined as
    $\norm{\bm{X}_{\cdot, t}} = \sum\limits_{n = 1}^N |\bm{X}_{n, t}|$, where $t \in \{1, \ldots, T\}$ is the frame index. The norm consistency loss is given by
    \begin{equation}
      \mathcal{L}_{norm} = \mathbb{E}_{\bm{X}, s} \left[\frac{1}{T}\sum\limits_{t =  1}^T\abs{\norm{\bm{X}_{\cdot, t}} - \norm{G(\bm{X}, s))_{\cdot, t}}}\right]
      \label{eq7}
    \end{equation}
    
    \noindent\textbf{Cycle consistency loss.} Lastly, we employ the cycle consistency loss \cite{zhu2017unpaired} to preserve all other features of the input
    \begin{equation}
      \mathcal{L}_{cyc} = \mathbb{E}_{\bm{X}, y_{src}, y_{trg}, s} \left[\norm{\bm{X} - G(G(\bm{X}, s), \tilde{s}))}_1\right]
      \label{eq8}
    \end{equation}
    where $\tilde{s} = S(\bm{X}, y_{src})$ is the estimated style code of the input in the source domain $y_{src} \in \mathcal{Y}$. 
    
    \noindent\textbf{Full objective.} Our full generator objective functions can be summarized as follows:
    \begin{equation}
    \begin{aligned}
        \min_{G, S, M} &\text{  }
       \mathcal{L}_{adv} + \lambda_{advcls}\mathcal{L}_{advcls} + \lambda_{sty}  \mathcal{L}_{sty} \\
        &-\lambda_{ds}  \mathcal{L}_{ds} + \lambda_{f0}  \mathcal{L}_{f0} + 
      \lambda_{asr}  \mathcal{L}_{asr} \\
      &+\lambda_{norm}  \mathcal{L}_{norm} +
      \lambda_{cyc}  \mathcal{L}_{cyc}
    \end{aligned}
      \label{eq9}
    \end{equation}
    where $\lambda_{advcls}, \lambda_{sty}, \lambda_{ds}, \lambda_{f0}, \lambda_{asr}, \lambda_{norm}$ and $ \lambda_{cyc}$ are hyperparameters for each term. 
    
    Our full discriminators objective is given by: 
    \begin{equation}
        \min_{C, D} -\mathcal{L}_{adv} + \lambda_{cls}\mathcal{L}_{cls} 
      \label{eq10}
    \end{equation}
    where $\lambda_{cls}$ is the hyperparameter for source classifier loss $\mathcal{L}_{cls}$, which is give by 
    \begin{equation}
      \mathcal{L}_{cls} = \mathbb{E}_{\bm{X}, y_{src}, s}\left[\textbf{CE}(C(G(\bm{X}, s)), y_{src})\right]
      \label{eq10}
    \end{equation}
\section{Experiments}

\subsection{Datasets }
    For a fair comparison, we train both the baseline model and our framework on the same 20 selected speakers reported in \cite{chou2018multi} from VCTK dataset \cite{yamagishi2019cstr}. To demonstrate the ability to convert to stylistic speech, we train our framework on 10 randomly selected speakers from the JVS dataset \cite{takamichi2019jvs} with both regular and falsetto utterances. We also train our model on 10 English speakers from the emotional speech dataset (ESD) \cite{zhou2020seen} with all five different emotions. We train our ASR and F0 model on the \textit{train-clean-100} subset from the LibriSpeech dataset \cite{panayotov2015librispeech}. All datasets are resampled to 24 kHz and randomly split according to an $80\%/10\%/10\%$ of train/val/test partitions. For the baseline model, the datasets are downsampled to 16 kHz. 
    
\subsection {Training details}
    We train our model for 150 epochs, with a batch size of 10 two-second long audio segments. We use AdamW \cite{loshchilov2017decoupled} optimizer with a learning rate of 0.0001 fixed throughout the training process. The source classifier joins the training process after 50 epochs. We set $\lambda_{cls} = 0.1, \lambda_{advcls} = 0.5, \lambda_{sty} = 1, \lambda_{ds} = 1, \lambda_{f0} = 5, \lambda_{asr} = 1, \lambda_{norm} = 1$ and $ \lambda_{cyc} = 1$. The F0 model is trained with pitch contours given by World vocoder \cite{morise2016world} for 100 epochs, and the ASR model is trained at phoneme level for 80 epochs with a character error rate (CER) of 8.53\%. 
    AUTO-VC is trained with one-hot embedding for 1M steps.
    
\subsection {Evaluations}
\begin{table}[t]
  \caption{Mean and standard error with subjective metrics. }
  \label{tab:table1}
  \centering
{
\begin{tabular}{cccccc}
\Xhline{2\arrayrulewidth}
\textbf{Method}                       & \textbf{Type} & \textbf{MOS}& \textbf{SIM}         \\ \hline
\multirow{2}{*}{Ground truth}              & M  & $4.64 \pm 0.14$   & \multirow{2}{*}{---}  \\
                             & F  &  $4.53 \pm 0.15$   &  \\                  \hline
\multirow{4}{*}{StarGANv2-VC} & M2M  & $4.02 \pm 0.22$   & \multirow{4}{*}{$3.86 \pm 0.24$}\\
                             & F2M  & $4.06 \pm 0.28$   &                                      \\
                             & F2F  & $4.25 \pm 0.26$   &                                          \\
                             & M2F  & $4.03 \pm 0.26$   &                                         \\ \hline
\multirow{4}{*}{AUTO-VC}     & M2M  & $2.70 \pm 0.26$    & \multirow{4}{*}{$3.57 \pm 0.27$} \\
                             & F2M  & $2.34 \pm 0.20$   &                                         \\
                             & F2F  & $2.86 \pm 0.23$   &                                          \\
                             & M2F  & $2.17 \pm 0.23$   &                      \\
                             
                             \Xhline{2\arrayrulewidth}

\end{tabular}}
\end{table}

    We evaluate our model with both subjective and objective metrics. The ablation study is conducted with only objective evaluations because the subjective evaluations are expensive and time-consuming. We use a pre-trained Parallel WaveGAN\footnote{Available at \url{https://github.com/kan-bayashi/ParallelWaveGAN}} \cite{yamamoto2020parallel} to synthesize waveforms from mel-spectrogram. We downsampled our waveforms to 16 kHz to match the sample rate of \cite{qian2019autovc}.
    
    \noindent\textbf{Subjective metric.} We randomly selected 5 male and 5 female speakers as our target speakers. The source speakers are randomly chosen from all 20 speakers to form 40 conversion pairs. Both source and ground truth samples were chosen to be at least 5-second long to ensure that there is enough information to judge the naturalness and similarity. We asked 46 online subjects to rate the naturalness of each audio clip on Amazon Mechanical Turk \footnote{The survey can be found at \url{https://survey.alchemer.com/s3/6266556/SoundQuality2}} on a scale of 1 to 5, where 1 indicates completely distorted and unnatural and 5 indicates no distortion and completely natural. Furthermore, we asked the subjects to rate from 1 to 5 whether the speakers of each pair of the audio clips could have been the same person, disregarding the distortion, speed, and tone of speech, where 1 indicates completely different speakers and 5 indicates exactly the same speaker. The subjects were not told whether an audio clip is the ground truth or converted. To make sure that the subjects did not complete the survey with random selections, we  included 6 completely distorted and unintelligible audios as attention checks. The raters are excluded from the analysis if more than two of these samples were rated above 2. After excluding bad raters, we ended up with 43 subjects for our analysis. All raters are self-identified as native English speakers and are located in the United States.
    
    \noindent\textbf{Objective metric.} We use predicted mean opinion score (pMOS) from MOSNet \cite{lo2019mosnet} to conduct the ablation study. Das et. al. \cite{das2020predictions} suggest automatic speaker verification (ASV) can be used as an objective metric for speaker similarity. For simplicity, similar to \cite{polyak2020unsupervised}, we only train an AlexNet \cite{krizhevsky2012imagenet} for speaker recognition on the 20 selected speakers and report the classification accuracy (CLS). We also report the character error rate (CER) using the aforementioned ASR model for intelligibility. 
    
\section{Results}
\begin{table}[t]
  \caption{Results with objective metrics. Full StarGANv2-VC represents that all loss functions are included. CLS for ground truth is evaluated on the test set of AlexNet. All other metrics are evaluated on 1,000 samples with random source and target pairs. Mean and standard deviation of pMOS are reported. }
  \label{tab:table2}
  \centering
\begin{tabular}{cccc}
\Xhline{2\arrayrulewidth}
\textbf{Method}   & \textbf{pMOS} & \textbf{CLS}  & \textbf{CER}  \\ \hline
Ground truth        &   $4.02 \pm 0.46$    &  98.67\%   &  11.47\%  \\\hline
Full StarGANv2-VC        &   $3.95 \pm 0.50$    &  96.20\%  & 12.55\% \\
$\lambda_{f0} = 0$          &   $3.99 \pm 0.54$     &  $96.50\%$ & $13.02\%$   \\
$\lambda_{asr} = 0$          &    $3.95 \pm 0.53$    &  96.90\%   &  30.34\% \\
$\lambda_{norm} = 0$         &     $3.83 \pm 0.47$   &   $96.30\%$ & 15.58\%  \\
$\lambda_{advcls} = 0$  &    $3.98 \pm 0.45$    &    $63.90\%$  & 12.33\% \\\hline
AUTO-VC         &    $3.43 \pm 0.51$    &  50.30\%  & 47.43\%
\\\Xhline{2\arrayrulewidth}
\end{tabular}
\end{table}
    The results of the survey show that our method significantly outperforms the AUTO-VC model in terms of rated naturalness in all four conversion types (MOS column in Table~\ref{tab:table1}, $p<0.001$ for randomization test, $p<0.001$ for t-test). 
    The survey results also indicate that converted samples using our framework are significantly more similar to the ground truth in terms of speakers' identity than the baseline model (SIM column in Table~\ref{tab:table1}). 
    The accuracy of the speaker recognition model on samples converted using our framework is much higher than those converted using AUTO-VC , and the predicted MOS of samples converted using our model is also significantly higher than AUTO-VC (see Table~\ref{tab:table2}). Lastly, CER on audio clips converted using our model is significantly lower than those converted using AUTO-VC. 
    
    The ablation study shows that $\lambda_{asr}$ is crucial for converting linguistic content, and removing $\lambda_{advcls}$ significantly decreases the speaker recognition accuracy. We notice that removing $\lambda_{norm}$ introduces noises to the converted audios during the silent intervals of the source audios. We also note that removing $\lambda_{f0}$ does not necessarily lower pMOS, CLS, or CER because all these three metrics are not sensitive to F0. However, we find that the converted samples have unnatural intonations by examining the converted samples without $\lambda_{f0}$. 
    
    Lastly, we show that our framework is generalizable to a variety of voice conversion tasks with audio demos. Demo audio samples can be found online\footnote{Available at \url{https://starganv2-vc.github.io/}}.
    
\section{Conclusion}
    We present an unsupervised framework with StarGAN v2 for voice conversion with novel adversarial and perceptual losses to achieve state-of-art performance in terms of both naturalness and similarity. In VCC2020, only PGG-VC based methods \cite{sun2016phonetic} could achieve MOS higher than 4.0 with autoregressive vocoders, and the model with highest MOS was based on TTS using a transformer architecture \cite{huang2019voice}. Our model achieves similar MOS but without the need of text labels during training and the use of autoregressive vocoders. Our framework can also learn to convert to stylistic speech such as emotional acting or falsetto speech from plain reading source speech. Furthermore, our model generalizes to various voice conversion tasks, such as any-to-many, cross-lingual and singing conversion, without the need for explicit training. Using Parallel WaveGAN vocoder, our model can convert an audio clip hundreds of times faster than real time on Tesla P100, which makes it suitable for real-time voice conversion applications. Future work includes improving the quality of any-to-many, cross-lingual, and singing conversion schemes with our framework. We would also like to develop a real-time voice conversion system with our models.  
    
\section{Acknowledgements}

We  would like to acknowledge Ryo Kato for proposing StarGAN v2 for voice conversion and funding is from the National Institute of Health, NIDCD. 
\newpage

\bibliographystyle{IEEEtran}

\bibliography{mybib}

\end{document}